# Single-photon upconversion

V. Yu. Shishkov[1,2], E. S. Andrianov[1,2], A. A. Pukhov[1,2,3], A. P. Vinogradov[1,2,3], and A. A. Lisyansky[4,5]

[1]Dukhov Research Institute for Automatics, 22 Sushchevskaya, Moscow 127055, Russia
[2]Moscow Institute of Physics and Technology, 9 Institutskiy per., 141700 Dolgoprudny, Moscow reg., Russia
[3]Institute for Theoretical and Applied Electromagnetics RAS, 13 Izhorskaya, Moscow 125412, Russia
[4]Department of Physics, Queens College of the City University of New York, Flushing, NY 11367, USA
[5]The Graduate Center of the City University of New York, New York, New York 10016, USA

The phenomenon of upconversion, in which a system sequentially absorbs two or more photons and emits a photon of a higher frequency, has been used in numerous applications. These include high-resolution non-destructive bioimaging[1-3], deep-penetrating photodynamic therapy[4,5], and photovoltaic devices[6,7]. Due to the multi-photon mechanism of upconversion, its quantum yield cannot exceed 50%. We propose a new mechanism of upconversion, which is based on single-photon absorption; in this process, unlike in multiple-photon upconversion, the quantum yield can be higher than 50%. We show that in a system of two atoms interacting with a reservoir, a low-frequency excitation of one atom can be upconverted into a high-frequency excitation of another atom. The energy required for such an upconversion is drawn from the reservoir, which destroys coherence. Decoherence leads to the transition of the system from the pure state with a small energy dispersion to the mixed state with greater dispersion of energy, while the system entropy increases. The phenomenon of single-photon upconversion can be used to increase the efficacy of devices utilizing upconversion.

**Introduction**

In 1958, Bloembergen suggested a sensor of low-frequency photons based on their conversion into a high-frequency photon[8]. In that sensor a three-level system with excitation energies from the ground state $\hbar\omega_{12}$ and $\hbar\omega_{13}$ is continuously illuminated with photons with the energy $\hbar\omega_{23}$. When the system consequently absorbs photons with frequencies $\omega_{12}$ and $\omega_{23}$, it emits a photon with the higher frequency $\omega_{13} = \omega_{12} + \omega_{23}$. Later[9], a number of systems absorbing several low-frequency photons and emitting a high-frequency photon were suggested (for review see Refs. [10-12]). The phenomenon has been called upconversion. Unlike a conventional luminescence, in which high-frequency radiation is transformed into low-frequency radiation, in upconversion, low-frequency radiation absorbed by a system is reradiated in a higher frequency range[13-15] with the absorption of at least two photons for each emitted photon. When the difference in frequencies between low and high-frequency photons increases, a larger number of photons is required for upconversion further decreasing the quantum yield of the process.

In the present paper, we suggest a new phenomenon of single-photon upconversion in which only one photon absorbed by a low-frequency atom is required for an excitation of a high-frequency atom. This phenomenon arises from the interaction of a quantum system with a reservoir that causes decoherence of the system. In the energy transfer from the low-frequency atom to the high-frequency one, the deficit of the energy is compensated by the energy inflow from the reservoir with positive temperature.

To observe single-photon upconversion, the decoherence time $\tau_{DC}$ should be much smaller than the lifetime of the excited states, $\tau_L$. For a number of dyes and biological molecules with $\tau_{DC} \sim 10^{-14}$ s and $\tau_L \sim 10^{-9}$ s, the condition $\tau_{DC} \ll \tau_L$ is easily realized[16-19]. We show that in the process of single-photon upconversion, the time $\tau_{UC}$ of the system excitation emerges as a new characteristic time-scale. Unlike multi-photon upconversion, the suggested mechanism can provide the quantum yield greater than 50% regardless of the difference in frequencies between low- and high-frequency photons.

**Results**

The dynamics of a system of two atoms interacting with a reservoir can be described by the density matrix which satisfies the Lindblad equation[20]

$$\frac{d\hat{\rho}(t)}{dt} = \hat{\mathcal{L}}[\hat{\rho}(t)] = -i[\hat{H}_0, \hat{\rho}(t)] + \sum_j \left( \hat{L}_j^+ \hat{\rho}(t) \hat{L}_j - \frac{1}{2} \hat{L}_j^+ \hat{L}_j \hat{\rho}(t) - \frac{1}{2} \hat{\rho}(t) \hat{L}_j^+ \hat{L}_j \right), \quad (1)$$

where $\hat{H}_0 = \omega_1 \hat{\sigma}_1^+ \hat{\sigma}_1 + \omega_2 \hat{\sigma}_2^+ \hat{\sigma}_2 + g(\hat{\sigma}_1^+ \hat{\sigma}_2 + \hat{\sigma}_1 \hat{\sigma}_2^+)$, $\omega_1$ and $\omega_2$ are excitation frequencies of two-level atoms, and $g$ is the interaction constant between these atoms, $\hat{\sigma}_1^+, \hat{\sigma}_1, \hat{\sigma}_2^+$, and $\hat{\sigma}_2$ are creation and annihilation operators in the first and second atoms, respectively. In Eq. (1), $\hbar = 1$ and the sum is taken over all dissipation channels with different Lindblad operators $\hat{L}_j$.

We begin with a simplified system. Since we assume that $\tau_{DC} \ll \tau_L$, we neglect the reservoir of vacuum photons and put $\tau_{L1} = \tau_{L2} = \infty$. We also assume that only the low-frequency atom interacts with the reservoir. Thus, $\hat{L}$, which only assures decoherence, can be chosen in the form $\hat{L} = \hat{\sigma}_1^+ \hat{\sigma}_1 / \sqrt{\tau_{DC}}$, where the constant $\tau_{DC}$ is the decoherence time[21]

$$\tau_{DC}^{-1} = \pi \int_0^{+\infty} d\omega \left[ v^2(\omega) |\alpha_\omega|^2 \bar{n}(\omega,T)(\bar{n}(\omega,T)+1) \right], \quad (2)$$

where $v(\omega)$ is the density of states and $\bar{n}(\omega,T)$ is an average number of excitations with the frequency $\omega$ in the reservoir which is determined by the temperature $T$.

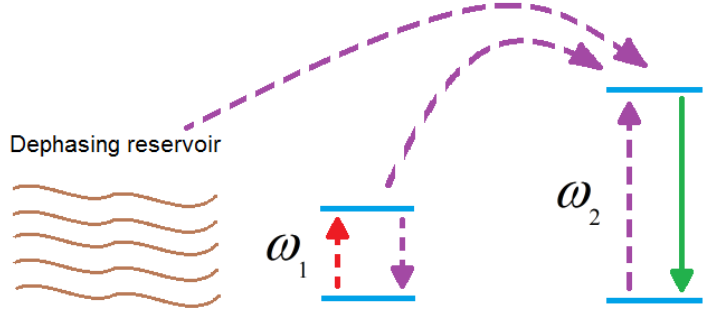

Fig. 1. Schematics of the energy levels in a system of two atoms.

Our analysis shows that if $\tau_{DC} \neq \infty$, Eq. (1) has three stationary solutions:

$$\hat{\rho}_1 = \begin{pmatrix} 1 & 0 & 0 & 0 \\ 0 & 0 & 0 & 0 \\ 0 & 0 & 0 & 0 \\ 0 & 0 & 0 & 0 \end{pmatrix}, \quad \hat{\rho}_2 = \begin{pmatrix} 0 & 0 & 0 & 0 \\ 0 & 0 & 0 & 0 \\ 0 & 0 & 0 & 0 \\ 0 & 0 & 0 & 1 \end{pmatrix}, \quad \hat{\rho}_3 = \begin{pmatrix} 0 & 0 & 0 & 0 \\ 0 & 0.5 & 0 & 0 \\ 0 & 0 & 0.5 & 0 \\ 0 & 0 & 0 & 0 \end{pmatrix}. \quad (3)$$

The solutions $\hat{\rho}_1$, $\hat{\rho}_2$, and $\hat{\rho}_3$, respectively, correspond to the states in which (1) both atoms are not excited, (2) both atoms are excited, and (3) probabilities to find atoms in excited states are the same. The final state of the system depends on its initial state. If initially, only the atom with the smaller transition frequency $\omega_1$ is excited, then in the final state, both atoms are excited with probability one-half. This is the case in which in the presence of decoherence, single-photon upconversion occurs: the low-frequency atom transfers excitation to the high-frequency atom. In the absence of the reservoir that causes decoherence, the system undergoes periodic non-resonant Forster transitions (the Rabi oscillations). In this case, the probability of the second atom being in the excited state is proportional to $g^2/(\omega_2 - \omega_1)^2 \ll 1$. Decoherence increases the probability of excitation of the high-frequency atom to one-half, as shown in our computer simulations (Fig. 2).

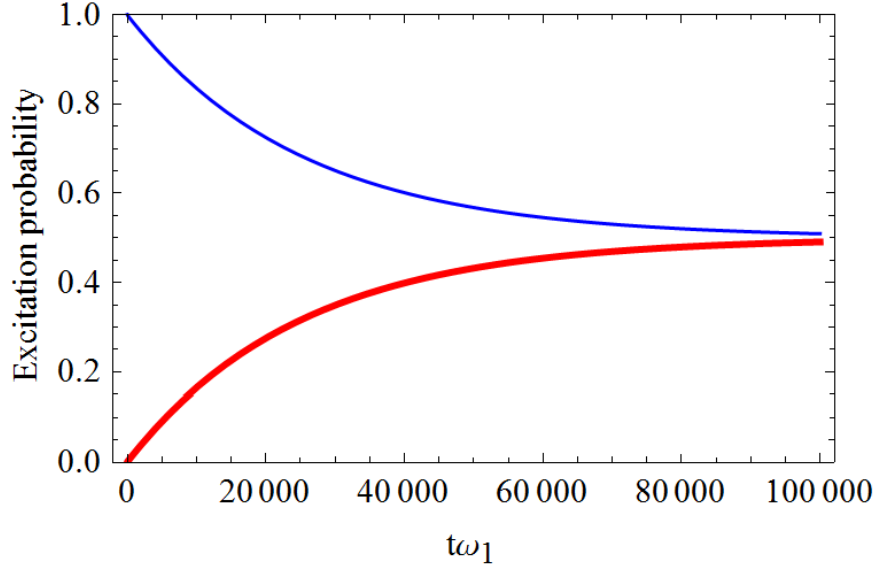

Fig. 2. The excitation probabilities. The probabilities of the atoms to be in the excited state are shown as functions of the dimensionless time $t\omega_1$ by the blue line (the first atom) and the red line (the second atom). These graphs are results of the numerical solution of Eq. (1) for the following values of the parameters: $\omega_2 = 2\omega_1$, $g/\omega_1 = 0.02$, and $\omega_1 \tau_{DC} = 20$.

As shown in Fig. 3, the total energy of two atoms, $\langle \hat{H}_0 \rangle$, increases when the excitation is transferred from the low-frequency atom to the high-frequency atom. Since the energy of an isolated system of two atoms is conserved, the only source of the increase is the energy influx from the reservoir. From a general point of view, such an energy transfer is allowed in a non-equilibrium system[22, 23]. Decoherence due to the interaction with the reservoir leads to the transition of the system from the pure state with small energy dispersion to the mixed state with greater dispersion (Fig. 3). Our computer simulation shows (Fig. 3) that the system entropy, $S = -Tr[\hat{\rho} \ln \hat{\rho}]$, increases, as it should.

Figure 3 shows that the time of the system upconversion emerges as a new characteristic time-scale $\tau_{UC}$. Assuming that the excitation probability of the high-frequency atom in Fig. 2 is fitted by the expression $(1 - \exp(-t/\tau_{UC}))/2$, the value of $\tau_{UC}$ can be extracted from the computer simulation data as $\omega_1 \tau_{UC} = -\omega_1 \left[ d \ln \left( d \langle \hat{\sigma}_2^+ \hat{\sigma}_2 \rangle / \omega_1 dt \right)/dt \right]^{-1} \approx 25000$. It means that $\tau_{UC} \gg \omega_1^{-1}, \omega_2^{-1}, g^{-1}, \tau_{DC}$ is the largest time in the system except for the times $\tau_{L1}$ and $\tau_{L2}$ that should remain the largest times in the problem as is assumed in our calculations. In the Methods section, the following expression for $\tau_{UC}$ is obtained

$$\tau_{UC} = \frac{(\omega_2 - \omega_1)^2}{2g^2} \tau_{DC}. \qquad (4)$$

Thus, the condition for the system behavior shown in Figs. 2 and 3 is $\tau_{L1}, \tau_{L2} \gg \tau_{UC} = (\omega_2 - \omega_1)^2 \tau_{DC} / 2g^2$. If $\tau_{UC}$ increases exceeding $\tau_{L1}$ and $\tau_{L2}$, the final probability of the high-frequency atom being in the excited state is no longer one half; it falls off approaching $g^2 / (\omega_2 - \omega_1)^2 \ll 1$.

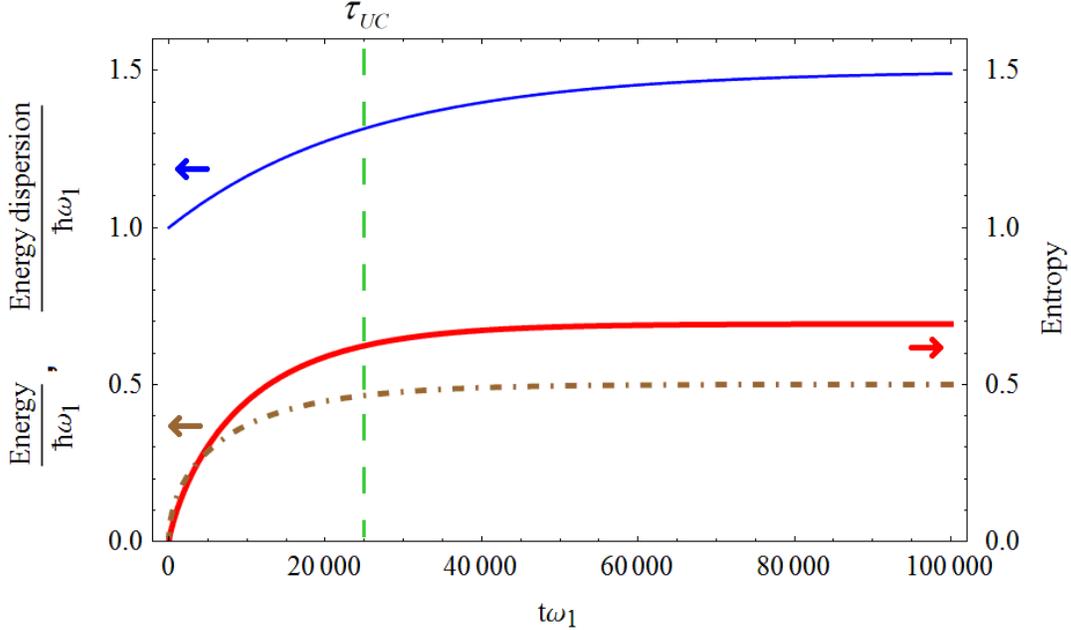

Fig. 3. The average energy (the solid thin blue line), energy dispersion (the dash-dotted brown line) and the von Neumann entropy (the solid thick red line) of two coupled atoms as functions of time. The graphs are calculated for the same values of the parameters as in Fig. 2.

So far, to describe single-photon upconversion, we have used a simplified model in which both $\tau_{L1}$ and $\tau_{L2}$ are infinite. The effect of the energy transfer from the reservoir to the coupled atoms is still preserved when these times are finite. In this case, however, the average energy shown in Fig. 3 is multiplied by the factor $e^{-\gamma t}$, where $\gamma \sim \tau_L^{-1}$. The predicted phenomenon can only occur if the rate of the longitudinal relaxation that results in the energy outflow from the system is smaller than the decoherence rate. In the next section, we present the results of computer simulation in which atom lifetimes are assumed to be finite and both atoms interact with the reservoir.

Note that there are no requirements for the reservoir temperature. The reservoir does not have to be "hot" for upconversion to occur. According to Eq. (2), its temperature can be smaller than all characteristic energies of the system. Its value only affects the decoherence time $\tau_{DC}$, and consequently, the upconversion time $\tau_{UC}$.

The quantum yield, *QY*, of upconversion is defined as the ratio of the number of low-frequency photons absorbed to the number of high-frequency photons emitted. If non-radiative decay is neglected, the *QY* can be obtained as (see Methods):

$$QY = \frac{1}{1 + \frac{\tau_{L2}}{\tau_{L1}} + 2\frac{\tau_{UC}}{\tau_{L1}}} \quad (5)$$

For $\tau_{L1} > \tau_{L2}, \tau_{UC}$, the quantum yield can be greater than 50% which is impossible for multi-photon upconversion. These inequalities are fulfilled for many systems including dye molecules and qubits.

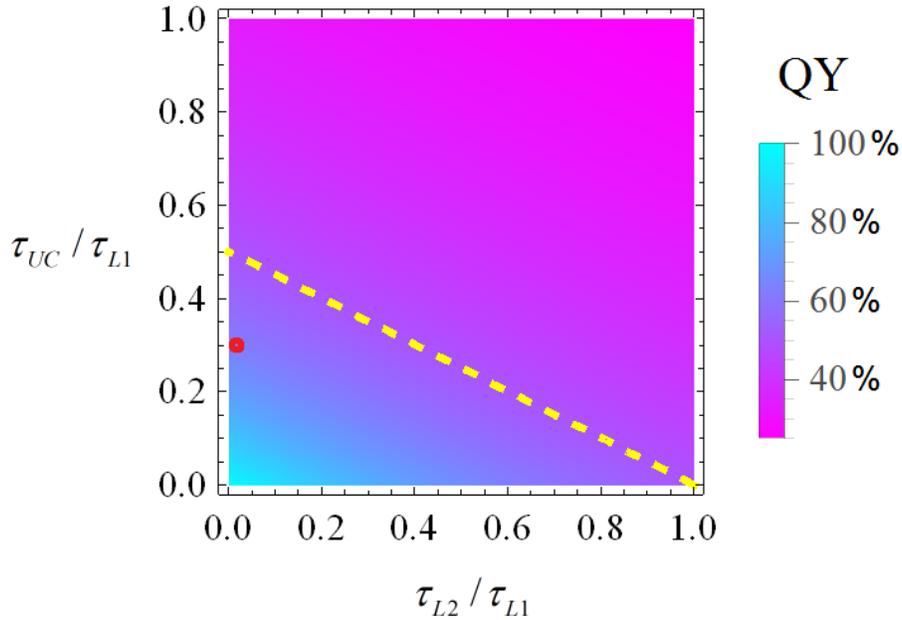

Fig. 4. Quantum yield as a function of $\tau_{L2}/\tau_{L1}$ and $\tau_{UC}/\tau_{L1}$. The yellow dashed line corresponds to the quantum yield of 50%. The red circle corresponds to the parameters for the colloidal quantum dots CdSe-ZnS and PbS described in the next section giving the quantum yield of 60%.

**Possible Experimental Verification**

Multiphoton upconversion in quantum dots has been observed in a number of experiments. These includes CdSe/NaYF4: Yb, Er nanoheterostructures[24], core-shell-shell PbSe/CdSe/CdS nanocrystals[25], carbon quantum dots[26], and colloidal double quantum dots[27]. With using quantum dots, upconversion opens up new opportunities for various applications. In particular, it allows for the conversion of near-infrared light to visible light. This is important for enhancement of the photocatalyst performance[26]. Also, for sensing and bioimaging, a near-infrared excitation is more usable because it is less harmful to biological objects[28]. However, due to the multiphoton nature of such upconversion, its quantum yield is low.

The effect considered in the paper can be verified experimentally with a system of two colloidal quantum dots (CQD's) which produces a quantum yield higher than 50%. Below, we consider a possible experiment, in which radiation intensities of systems of CQD's CdSe-ZnS (core radius 3 nm, shell thickness 5 nm), PbS (radius 1.6 nm), and their mixture are measured. The concentration of each CQDs is assumed to be $10^{18}\,\text{cm}^{-3}$. The wavelength of the pumping laser should coincide with the luminescence wavelength of the PbS CQD ($\lambda_1 = 1000$ nm). The intensities of luminescence of PbS, CdSe-ZnS, and their mixture are measured at the luminescence wavelength of CdSe-ZnS CQD ($\lambda_2 = 560$ nm).

For PbS, at the transition wavelength $\lambda_1 = 1000$ nm, longitudinal and transverse relaxation times are $\tau_{DC1} = 8$ fs and $\tau_{L1} = 1200$ ns, respectively[29, 30]. For CdSe-ZnS, for $\lambda_2 = 560$ nm these parameters (in the same order) are $\tau_{DC2} = 9$ fs and $\tau_{L2} = 4.8$ ns[31, 32]. The dipole moments of the transitions are $d_1 = 20$ D and $d_2 = 8$ D for PbS and CdSe-ZnS, respectively[33, 34].

The dynamics of the system is studied via computer simulation of the Lindblad equation (1) that for this system has the form:

$$\frac{d\hat{\rho}(t)}{dt} = \hat{\mathcal{L}}\left[\hat{\rho}(t)\right] = -i\left[\hat{H}_0 + \hat{H}_{ex}(t), \hat{\rho}(t)\right] + \hat{L}_{DC1}^+ \hat{\rho}(t)\hat{L}_{DC1} - \frac{1}{2}\hat{L}_{DC1}^+ \hat{L}_{DC1}\hat{\rho}(t) - \frac{1}{2}\hat{\rho}(t)\hat{L}_{DC1}^+ \hat{L}_{DC1},$$

$$+ \hat{L}_{DC2}^+ \hat{\rho}(t)\hat{L}_{DC2} - \frac{1}{2}\hat{L}_{DC2}^+ \hat{L}_{DC2}\hat{\rho}(t) - \frac{1}{2}\hat{\rho}(t)\hat{L}_{DC2}^+ \hat{L}_{DC2} + \hat{L}_{L1}^+ \hat{\rho}(t)\hat{L}_{L1} - \frac{1}{2}\hat{L}_{L1}^+ \hat{L}_{L1}\hat{\rho}(t) \qquad (6)$$

$$- \frac{1}{2}\hat{\rho}(t)\hat{L}_{L1}^+ \hat{L}_{L1} + \hat{L}_{L2}^+ \hat{\rho}(t)\hat{L}_{L2} - \frac{1}{2}\hat{L}_{L2}^+ \hat{L}_{L2}\hat{\rho}(t) - \frac{1}{2}\hat{\rho}(t)\hat{L}_{L2}^+ \hat{L}_{L2}$$

where subscripts 1 and 2 denote operators related to PbS and CdSe-ZnS CQD's, respectively, $\hat{H}_0 = \omega_1 \hat{\sigma}_1^+ \hat{\sigma}_1 + \omega_2 \hat{\sigma}_2^+ \hat{\sigma}_2 + g\left(\hat{\sigma}_1^+ \hat{\sigma}_2 + \hat{\sigma}_1 \hat{\sigma}_2^+\right)$. Pumping is described by the operator $\hat{H}_{ex} = \Omega_R \left(\hat{\sigma}_1^+ + \hat{\sigma}_1 + \hat{\sigma}_2^+ + \hat{\sigma}_2\right) \sin(\omega_1 t)$, where $\Omega_R$ is the Rabi constant which square is proportional to the intensity of the laser pumping. The Lindblad operators $\hat{L}_{DC1} = \hat{\sigma}_1^+ \hat{\sigma}_1 / \sqrt{\tau_{DC1}}$ and $\hat{L}_{DC2} = \hat{\sigma}_2^+ \hat{\sigma}_2 / \sqrt{\tau_{DC2}}$ describe decoherence and $\hat{L}_{L1} = \hat{\sigma}_1 / \sqrt{\tau_{L1}}$ and $\hat{L}_{L2} = \hat{\sigma}_2 / \sqrt{\tau_{L2}}$ describe radiation decay into the vacuum.

The analytical solutions to Eq. (6) that describe the stationary state for the atoms in the first and second samples are well known[35]. The stationary probabilities to find the PbS and CdSe-ZnS CQD's in the excited states are equal to $\langle \hat{\sigma}_1^+ \hat{\sigma}_1 \rangle = \tau_{L1} \tau_{DC1} \Omega_R^2$ and $\langle \hat{\sigma}_2^+ \hat{\sigma}_2 \rangle = \tau_{L2} \omega_2 \Omega_R^2 / (\omega_2^2 - \omega_1^2)$, respectively. To find the similar probability for the mixture of CQD's, we solve Eq. (6) numerically.

The probabilities obtained allow one to determine the total power radiated by each sample $W_j = \omega_j \langle \hat{\sigma}_j^+ \hat{\sigma}_j \rangle / \tau_{Lj}$ ($j = 1, 2$). Assuming that luminescent spectra have the Lorentz shape centered at the wavelengths of $\lambda_1 = 1000$ nm and $\lambda_2 = 560$ nm for PbS and CdSe-ZnS, respectively, we have

$$\frac{dW_j}{d\omega} = \omega_j \langle \hat{\sigma}_j^+ \hat{\sigma}_j \rangle \frac{1}{\pi \tau_{Lj} \tau_{DCj}} \frac{1}{\left(\omega - \omega_j\right)^2 + \tau_{DCj}^{-2}},$$

where $\omega_j = 2\pi c / \lambda_j$.

In Fig. 5, the spectral power densities, $dW_j/d\omega$, at the wavelength $\lambda_2 = 560$ nm as functions of the pump intensity are shown for the three samples described above. The linearity of the graphs indicates a single-photon character of the effect.

The linewidth of PbS CQD's is finite. Therefore, the luminescence illumination of these CQD's with light at the wavelength $\lambda_1 = 1000$ nm results in the luminescence response at the wavelength $\lambda_2 = 560$ nm (Fig. 5). Our calculations show that at the wavelength $\lambda_2 = 560$ nm, the first two samples together emit less light than the third sample (Figs. 5 and 6). The reason for this is that besides the usual luminescence from the PbS and CdSe-ZnS CQD's, there is an additional contribution due to single-photon upconversion in the third sample.

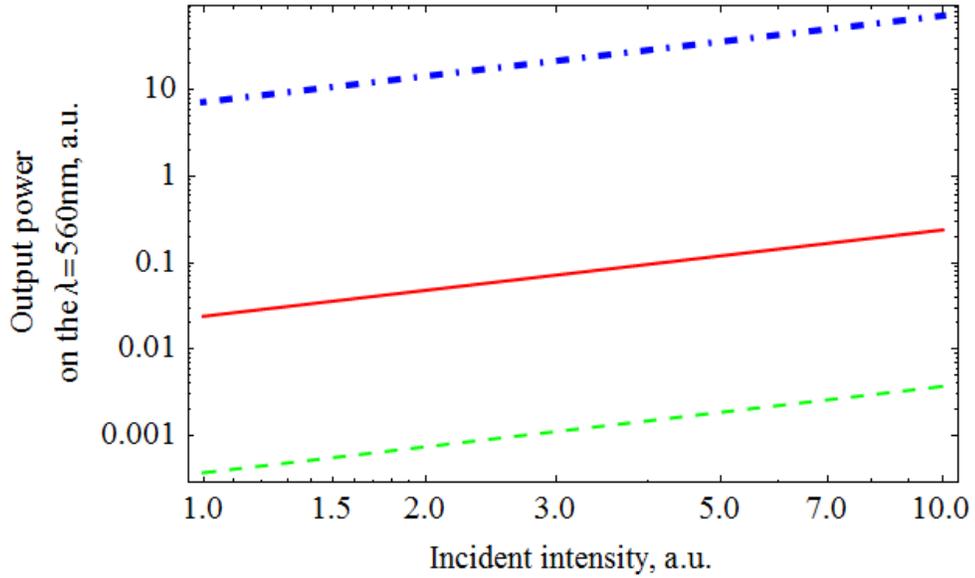

Fig. 5. The dependence of the spectral density on the power of high-frequency radiation on the pump intensity. The spectral densities for PbS, CdSe-ZnS, and the mixture of both CQD's are shown in the double logarithmic scale by solid red, dashed green, and dash-dotted blue lines, respectively.

For these calculations, we use parameters for standard CQD's for which inequalities $\tau_{L1}, \tau_{L2} \gg \tau_{DC1}, \tau_{DC2}$ required for single photon upconversion are easily fulfilled ($\tau_{L1} = 1200$ ns, $\tau_{DC1} = 8$ fs, $\tau_{L2} = 4.8$ ns, and $\tau_{DC2} = 9$ ns). For their mixture $\tau_{UC} \approx 380$ ns is smaller than $\tau_{L1}$. Therefore, for this system, the probability of single-photon upconversion can be expected to be high. Numerical calculations show that the probability of CdSe-ZnS CQD's upconversion is about 60%.

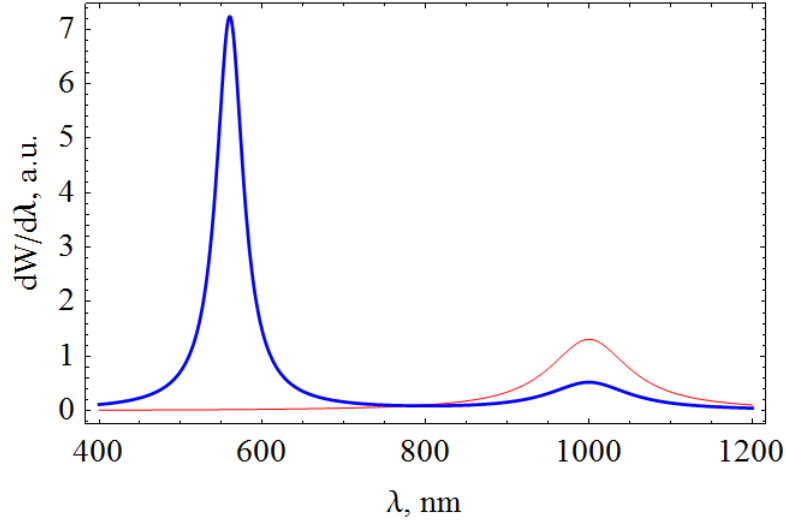

Fig. 6. The dependence of the spectral power density on the frequency. The sum of spectral power densities of PbS and CdSe-ZnS CQD's are shown by the thin red line, and the spectral power density of their mixture is shown by the thick blue line. The right and left peaks correspond to the wavelengths $\lambda_1 = 1000$ nm and $\lambda_2 = 560$ nm, respectively. All the systems are pumped at $\lambda_1 = 1000$ nm.

**Conclusion**

The suggested mechanism of single-photon upconversion is based on a counterintuitive phenomenon of pumping energy out of a thermal reservoir whose temperature can be much smaller than the energy of the emitted photon. The direction of the energy transfer only depends on the ratio of probabilities of initial excitations of the atoms: if the probability of the excitation of the low-frequency atom is higher than that of the high-frequency atom, then the energy is transferred from the reservoir, and vice versa.

We note the universality of the phenomenon discussed in the paper. When the initial state of the system is fixed, the final state depends neither on the ratio of transition frequencies of the atoms nor on constants of decoherence and coupling between atoms.

**Methods**

*Equations for expected values*

In the Schrödinger representation, in which the density matrix operator is the only operator that depends on time, we obtain the system of equations for operator averages[36] $\hat{\sigma}_1^+ \hat{\sigma}_1$, $\hat{\sigma}_2^+ \hat{\sigma}_2$, and $\hat{\sigma}_1^+ \hat{\sigma}_2$:

$$\frac{d}{dt}\langle \hat{\sigma}_1^+ \hat{\sigma}_1 \rangle = ig\left(\langle \hat{\sigma}_1 \hat{\sigma}_2^+ \rangle - \langle \hat{\sigma}_1^+ \hat{\sigma}_2 \rangle\right), \quad (7)$$

$$\frac{d}{dt}\langle \hat{\sigma}_2^+ \hat{\sigma}_2 \rangle = -ig\left(\langle \hat{\sigma}_1 \hat{\sigma}_2^+ \rangle - \langle \hat{\sigma}_1^+ \hat{\sigma}_2 \rangle\right), \quad (8)$$

$$\frac{d}{dt}\langle\hat{\sigma}_1^+\hat{\sigma}_2\rangle = \left(-i(\omega_2-\omega_1)-\frac{1}{2\tau_{DC}}\right)\langle\hat{\sigma}_1^+\hat{\sigma}_2\rangle - ig\left(\langle\hat{\sigma}_1^+\hat{\sigma}_1\rangle-\langle\hat{\sigma}_2^+\hat{\sigma}_2\rangle\right). \tag{9}$$

System (7)-(9) is closed and does not require further decoupling of correlators.

To find the upconversion time $\tau_{UC}$, we exclude correlators $\langle\hat{\sigma}_1^+\hat{\sigma}_1\rangle$ and $\langle\hat{\sigma}_2^+\hat{\sigma}_2\rangle$ from system (7)-(9) and introduce variables $x=\langle\hat{\sigma}_1^+\hat{\sigma}_2+\hat{\sigma}_2^+\hat{\sigma}_1\rangle/2$ and $y=\langle\hat{\sigma}_1^+\hat{\sigma}_2-\hat{\sigma}_2^+\hat{\sigma}_1\rangle/2i$. As a result, we obtain a system of equations

$$\frac{d}{dt}x = -\frac{1}{2\tau_{DC}}x+(\omega_2-\omega_1)y, \tag{10}$$

$$\frac{d^2}{dt^2}y = -(\omega_2-\omega_1)\frac{d}{dt}x - \frac{1}{2\tau_{DC}}\frac{d}{dt}y + 2g^2 y. \tag{11}$$

The decay rates $\lambda$ are defined as eigenvalues of this linear system:

$$\lambda^3 + \lambda^2 \frac{1}{\tau_{DC}} + \lambda\left((\omega_2-\omega_1)^2 + 4g^2 + \frac{1}{4\tau_{DC}^2}\right) + \frac{2}{\tau_{DC}}g^2 = 0. \tag{12}$$

When $|\omega_2-\omega_1| \gg g \sim \tau_{DC}^{-1}$, Eq. (12) gives the attenuation rate of the slowest decaying solution. This rate is $\lambda \approx 2g^2\tau_{DC}^{-1}(\omega_2-\omega_1)^{-2}$ which corresponds to the inverse time of the upconversion (4).

*Rate equations*

To estimate the quantum yield, we need to consider that eventually the system emits a photon. For this purpose, Eq. (1) should be modified by adding corresponding terms to the Lindblad operator. The same result can be obtained by using the rate equations. The latter approach describes the system behavior more clearly.

Let us assume that at the initial moment our system absorbs a photon with the frequency $\omega_1$. As a result, the low-frequency atom is excited and the other one remains in the ground state. The further evolution of the system can be described by the rate equations[37, 38]

$$\frac{dN_1(t)}{dt} = -\frac{2}{\tau_{UC}}(N_1(t)-N_2(t)) - \frac{1}{\tau_{L1}}N_1(t),$$

$$\frac{dN_2(t)}{dt} = -\frac{2}{\tau_{UC}}(N_2(t)-N_1(t)) - \frac{1}{\tau_{L2}}N_2(t), \tag{13}$$

where $N_1 = \langle\hat{\sigma}_1^+\hat{\sigma}_1\rangle$ and $N_2 = \langle\hat{\sigma}_2^+\hat{\sigma}_2\rangle$ are probabilities to find the respective atom in the excited state. The initial conditions for Eqs. (13) are $N_1(0)=1$ and $N_2(0)=0$. Since we assume that at the initial moment only one photon is absorbed, the quantum yield is defined as

$$QY = \frac{1}{\tau_{L2}}\int_0^{+\infty} N_2(t)\,dt. \tag{14}$$

System (13) has an analytical solution, which defines the probability of the excitation of the low-frequency atom:

$$N_2(t) = \frac{\tau_{UC}^{-1}}{2\sqrt{\tau_{UC}^{-2} + (\tau_{L1}^{-1} - \tau_{L2}^{-1})^2}} \left(e^{-\lambda_- t} - e^{-\lambda_+ t}\right), \quad (15)$$

where

$$\lambda_{\pm} = \frac{1}{2}\left(\tau_{UC}^{-1} + \tau_{L1}^{-1} + \tau_{L2}^{-1} \pm \sqrt{\tau_{UC}^{-2} + (\tau_{L1}^{-1} - \tau_{L2}^{-1})^2}\right). \quad (16)$$

Equations (14)-(16) give the expression (5) for the quantum yield of upconversion.

**Acknowledgements**

A.A.L. would like to acknowledge support from the NSF under Grant No. DMR-1312707.